# WAKE EXCITATION OF VOLUME AND SURFACE POLARITONS BY A RELATIVISTIC ELECTRON BUNCH IN AN ION DIELECTRIC

*V.A. Balakirev, I.N. Onishchenko*

*National Science Center "Kharkov Institute of Physics and Technology", Kharkov, Ukraine*
*E-mail: onish@kipt.kharkov.ua*

The process of wake fields excitation by a relativistic electron bunch in ion dielectric media is considered. The space-time structure of the excited wakefield in an ion dielectric waveguide is investigated. It is shown that the wake field in the infrared (terahertz) ranges consists of the field of longitudinal polaritons and Cherenkov radiation, which is a set of electromagnetic eigen waves (transverse polaritons) of a dielectric waveguide. A branch of surface polaritons appears in a dielectric waveguide with an axial vacuum channel. The process of wakefield excitation of surface polaritons by a relativistic electron bunch is also considered.

## INTRODUCTION

The Cherenkov radiation effect of charged particles (bunches of charged particles) moving in a dielectric medium [1] can be used for realization of the wakefield method of charged particles acceleration [2-11]. Moreover, as a rule, the wakefield excitation process was considered without taking into account the frequency dispersion of the dielectric constant of the medium. Meanwhile, taking into account the dependence of the dielectric constant on frequency leads to a number of qualitative features of the picture of wakefield excitation in dielectric structures.

In the infrared frequency range, ion dielectrics, for example, alkali-halogen group, have three branches of electromagnetic waves [12-14]. High-frequency and low-frequency electromagnetic branches correspond to transverse electromagnetic waves. The third branch oscillations describes purely potential longitudinal oscillations of a dielectric medium with ion bonding. The frequency of potential waves (longitudinal optical phonons) is the zero of the dielectric constant $\varepsilon(\omega) = 0$. All of these branches of the oscillations of the dielectric – polaritons [15,16] will be excited during the propagation of a laser pulse [17,18] in ion dielectrics. In the present work, the process of excitation of wake electromagnetic fields in an ion dielectric medium by a relativistic electron bunch is investigated. For definiteness, we will consider ion alkali-halide dielectrics, which have a cubic crystal structure. The choice of alkaline-halide crystal dielectrics is primarily due to their relatively simple internal structure and, accordingly, a simple dependence of the dielectric constant $\varepsilon(\omega)$, which allows us to study the process of wakefields excitation by relativistic electron bunch by analytical methods. Our aim is to study wake wave intensity, the frequency spectrum of excited wakefields and their space-time structures.

To transport relativistic electron bunches, it is necessary to have a vacuum channel in the dielectric. On the other hand, in the presence of a dielectric-vacuum boundary in an ion dielectric, an additional branch of oscillations arises - surface polaritons [19-24]. Surface polaritons exist in the frequency range where the dielectric constant is negative. The wake excitation of surface polaritons (surface optical phonons) is also given attention in this work.

We restrict ourselves to the study of wake fields in the infrared and lower frequency ranges. This is due to the fact that for efficient excitation of the wake field by an electron bunch, it is necessary to achieve coherence of electromagnetic waves excitation. For this, it is necessary that the longitudinal and transverse dimensions of the electron bunch should be less (substantially less) than the length of the excited wave. For optical and especially ultraviolet frequency ranges, this requirement is very problematic. And if this requirement is not satisfied, the amplitude of the wake wave will be insignificant.

## 1. EXCITATION OF VOLUME POLARITONS BY A RELATIVISTIC ELECTRON BUNCH IN A DIELECTRIC WAVEGUIDE

Let's consider the homogeneous dielectric cylinder of radius $b$, the side surface of which is covered with a perfectly conductive metal film. Along the axis of the dielectric waveguide, an axisymmetric REB moves uniformly and rectilinearly. The initial system of equations contains Maxwell's equations

$$rot\vec{E} = -\frac{1}{c}\frac{\partial \vec{H}}{\partial t}, \quad rot\vec{H} = \frac{1}{c}\frac{\partial \vec{D}}{\partial t} + \frac{4\pi}{c}\vec{j}_b$$
$$div\vec{D} = 4\pi\rho, \quad div\vec{H} = 0, \qquad (1)$$

$\rho_b, \vec{j}_b$ are charge density and current of an electron bunch, $\vec{D} = \hat{\varepsilon}\vec{E}$ is electric displacement field, $\hat{\varepsilon}$ is dielectric constant operator of an ion dielectric.

The system of Maxwell equations (1) describes the excitation of an electromagnetic field by external charges and currents in a condensed dielectric medium.

### 1.1. DETERMINATION OF THE GREEN FUNCTION

We will solve the problem of wake field excitation by an axisymmetric electron bunch in a dielectric waveguide as follows. First, we determine the field (Green's function) of a moving charge in the form of an infinitely thin ring with a charge density

$$d\rho = -dQ\frac{1}{v_0}\frac{\delta(r-r_0)}{2\pi r_0}\delta(t-\frac{z}{v_0}-t_0), \qquad (2)$$

where $r$ is radial coordinate, $r_0$ is ring radius, $t_0$ is time of entry of an elementary ring bunch into the waveguide, $v_0$ is bunch velocity, $dQ(r_0,t_0)$ is the charge of the elementary ring connected with the current density of the bunch at the entrance to the dielectric waveguide $(z=0)$ $j_0(t_0,r_0)$ by the relation

$$dQ = j_0(t_0,r_0)2\pi r_0 dr_0 dt_0. \quad (3)$$

The current density of an elementary ring charge is determined by the expression

$$d\vec{j} = v_0 d\rho \vec{e}_z, \quad (4)$$

$\vec{e}_z$ is unit vector in longitudinal direction.

Let's consider a bunch of electrons with the current density

$$j_0(r_0,t_0) = j_0 R(r_0/r_b) T(t_0/t_b), \quad (5)$$

where the function $R(r_0/r_b)$ describes dependence of the bunch density on radius (transverse profile), $r_b$ is characteristic transverse bunch size, function $T(t_0/t_b)$ describes the longitudinal density profile of a bunch, $t_b$ is characteristic duration of the bunch. The value $j_0$ is connected with a full charge $Q$ by the relation $j_0 = Q/(s_{eff} t_{eff})$, where $s_{eff}$ is effective cross section of the bunch

$$s_{eff} = \pi r_b^2 \hat{\sigma}, \quad \hat{\sigma} = 2\int_0^\infty R(\rho)\rho d\rho,$$

and $t_{eff}$ is effective bunch duration

$$t_{eff} = \hat{\tau} t_b, \quad \hat{\tau} = 2\int_0^\infty T(\tau_0) d\tau_0.$$

Let us define the electromagnetic field excited by an elementary ring charge (2) and current (4) as

$$\vec{E}_G(r,r_0,z,t-t_0) = dQ\vec{E}(r,r_0,z,t-t_0), \quad (6)$$

where $\vec{E}(r,r_0,z,t-t_0)$ is the electric field excited by an ring bunch with a unit charge. Then the full electromagnetic field, excited by an electron bunch of finite dimensions, is found by summing (integrating) the fields of elementary ring bunches.

$$\vec{E}(r,z,t) = \int_0^b 2\pi r_0 dr_0 \int_{-\infty}^t dt_0 j(r_0,t_0)\vec{E}(r,r_0,z,t-t_0). \quad (7)$$

Taking into account relations (3), (5), (7), this expression can be written as follows

$$\vec{E}(r,z,t) = \frac{2\pi Q}{s_{eff} t_{eff}} \int_0^b R\left(\frac{r_0}{r_b}\right) r_0 dr_0 \times$$

$$\times \int_{-\infty}^t T\left(\frac{t_0}{t_b}\right) \vec{E}(r,r_0,z,t-t_0) dt_0. \quad (8)$$

The next step in solution of the problem is determination the electromagnetic field (Green's function) (6) of an elementary ring electron bunch.

We expand the quantities in the Maxwell equations (1) into the Fourier integrals over frequencies

$$\left(\vec{E}_G(\bar{t}), \vec{H}_G(\bar{t})\right) = \int_{-\infty}^\infty \left(\vec{E}_{G\omega}, \vec{H}_{G\omega}\right) e^{-i\omega\bar{t}} d\omega, \quad (9)$$

$$\left(d\vec{j}_b, d\rho_b\right) = -dQ \frac{\delta(r-r_0)}{2\pi r_0} \frac{1}{2\pi} \int_{-\infty}^\infty \left(\vec{e}_z, \frac{1}{v_0}\right) e^{-i\omega\bar{t}} d\omega, \quad (10)$$

$\bar{t} = \tau - t_0$, $\tau = t - z/v_0$, $\vec{E}_G, \vec{H}_G$ is electromagnetic field (Green's function) (6) of elementary current and charge.

The system of Maxwell equations (1), taking into account relations (2), (4), can be transformed to the equation for the longitudinal Fourier component of the electric field

$$\frac{1}{r}\frac{d}{dr} r \frac{dE_{Gz\omega}}{dr} + k_\perp^2 E_{Gz\omega} = \frac{i}{\pi} \frac{k_\perp^2}{\omega\varepsilon(\omega)} dQ \frac{\delta(r-r_0)}{r_0}, \quad (11)$$

$$k_\perp^2 = k_0^2 \varepsilon(\omega) - k_l^2, \quad k_l = \omega/v_0, \quad k_0 = \omega/c,$$

$\varepsilon(\omega)$ is dielectric constant of an ion dielectric. On the perfectly conducting side surface of the dielectric waveguide $r = b$, the longitudinal component of the electric field vanishes

$$E_{Gz\omega}(r=b) = 0.$$

We will search the longitudinal Fourier component of the electric field $E_{Gz\omega}$ in the form of a series of Bessel functions

$$E_{Gz\omega} = \sum_{n=1}^\infty C_n(\omega) J_0(\lambda_n r/b), \quad (12)$$

where $\lambda_n$ are the roots of the Bessel function $J_0(x)$. Using the orthogonality of the Bessel functions $J_0(\lambda_n r/b)$, from equation (11) we find the expansion coefficients

$$C_n(\omega) = \frac{idQ}{\pi\omega\varepsilon(\omega)} \frac{k_\perp^2}{k_\perp^2 - \frac{\lambda_n^2}{b^2}} \frac{J_0(\lambda_n r_0/b) J_0(\lambda_n r/b)}{N_n}. \quad (13)$$

Here $N_n = \frac{b^2}{2} J_1^2(\lambda_n)$ is wave norm. Accordingly, for the longitudinal component of the electric field we have the following expression

$$E_{Gz}(r,\bar{t}) = \frac{2i}{\pi b^2} dQ \sum_{n=1}^\infty \frac{J_0(\lambda_n r_0/b) J_0(\lambda_n r/b)}{J_1^2(\lambda_n)} S_n(\bar{t}), \quad (14)$$

where

$$S_n(\bar{t}) = \int_{-\infty}^\infty e^{-i\omega\bar{t}} \frac{d\omega}{\omega} \frac{k_\perp^2(\omega)}{\varepsilon(\omega) D_n(\omega)}, \quad (15)$$

$$D_n(\omega) = k_0^2 \varepsilon(\omega) - k_l^2 - \frac{\lambda_n^2}{b^2}. \quad (16)$$

For the infrared range, the expression for the dielectric constant takes the form [12-14]

$$\varepsilon(\omega) = \varepsilon_{opt} \frac{\omega^2 - \omega_L^2}{\omega^2 - \omega_T^2}, \quad (17)$$

$\varepsilon_{opt}$ is optical dielectric permittivity. Frequency $\omega_L$ is zero of dielectric permittivity $\varepsilon(\omega_L) \equiv 0$. This frequency is the frequency of longitudinal optical phonons and lies in the infrared frequency range. We also note that the frequency $\omega_L$ are also cutoff frequencies for normal incidence of electromagnetic waves on a plane dielectric layer. In turn, the frequency $\omega_T$ is the pole of the dielectric constant and determines the absorption line of the electromagnetic waves of the ion crystal in the infrared frequency range. In the vicinity of this frequency, the imaginary part of the dielectric constant and, accordingly, the energy loss of electromagnetic waves is increased abnormally. The frequency $\omega_T$ is the frequency of transverse optical phonons. We note that the optical longitudinal and transverse acoustic branches of the oscillations are characterized by the fact that in the unit cell of the ion crystal, oppositely charged ions are shifted towards each other. In this case, the center of mass of the unit cell remains motionless. Like the frequency of longitudinal optical phonons, the frequency of transverse optical phonons lies in the infrared range. Note also that from the above relation for the permittivity (17) follows from the well-known Lyddane-Sachs-Teller relation [12-14]

$$\frac{\omega_L^2}{\omega_T^2} = \frac{\varepsilon_{st}}{\varepsilon_{opt}},$$

which determines the relationship between the frequencies of longitudinal and transverse phonons through the values of static $\varepsilon_{st}$ and optical $\varepsilon_{opt}$ permittivity.

The zeros of the dielectric constant are the poles of the integrand (15). Calculating the residues at the poles $\omega = \pm \omega_L - i0,$ we find the potential part of the Green's function

$$E_{Gz}^{(l)}(r,\bar{t}) = 2dQ \frac{k_L^2}{\varepsilon_{eff}} G(k_L r, k_L r_0) \vartheta(\bar{t}) \cos \omega_L \bar{t}, \quad (18)$$

$$k_L = \omega_L / v_0, \quad \varepsilon_{eff} = \frac{\varepsilon_{st}\varepsilon_{opt}}{\Delta\varepsilon}, \quad \varepsilon_{st} = \varepsilon_{opt} \frac{\omega_L^2}{\omega_T^2},$$

$$G(k_\alpha r, k_\alpha r_0) = \begin{cases} \frac{I_0(k_\alpha r_0)}{I_0(k_\alpha b)} \Delta_0(k_\alpha r, k_\alpha b), \ r > r_0, \\ \frac{I_0(k_\alpha r)}{I_0(k_\alpha b)} \Delta_0(k_\alpha r_0, k_\alpha b), \ r < r_0, \end{cases}$$

$$\Delta_0(k_\alpha r, k_\alpha b) = I_0(k_\alpha b) K_0(k_\alpha r) - I_0(k_\alpha r) K_0(k_\alpha b),$$

$$k_\alpha = k_L.$$

The integrand also has poles that are the roots of the equation

$$D_n(\omega) = \frac{\omega^2}{c^2} \varepsilon(\omega) - \frac{\omega^2}{v_0^2} - \frac{\lambda_n^2}{b^2} = 0. \quad (19)$$

Equation (19) determines the frequency spectrum of the radial harmonic with the number $n$ of electromagnetic waves excited by the relativistic electron bunch in ion dielectric waveguide. With respect to the square of the frequency $\omega^2$, the spectrum equation (19) reduces to determining the roots of the quadratic equation. The frequencies $\omega_{phn}^{(\mp)}$, corresponding to the roots of this equation, lie in the microwave and infrared ranges. The spectrum equation (19) can be written as follows

$$\omega^2 \left( \varepsilon_{opt} \frac{(\omega^2 - \omega_L^2)}{(\omega^2 - \omega_T^2)} - \frac{1}{\beta_0^2} \right) = \omega_n^2,$$

where $\beta_0 = v_0/c$, $\omega_n = \lambda_n c / b$ is cutoff frequency of a vacuum waveguide of radius $b$. The roots of this equation are of the form

$$\omega_{phn}^{(\mp)2} = \frac{1}{2} \left( \omega_{Gn}^2 \mp \sqrt{\omega_{Gn}^4 - 4\omega_{gn}^4} \right). \quad (20)$$

Here

$$\omega_{Gn}^2 = \frac{1}{d_{opt}} \left( \omega_T^2 d_{st} + \omega_n^2 \right), \ \omega_{gn}^4 = \frac{1}{d_{opt}} \omega_T^2 \omega_n^2,$$

$$d_{opt} = \varepsilon_{opt} - \beta_0^{-2}, \ d_{st} = \varepsilon_{st} - \beta_0^{-2}.$$

For the frequency $\omega_{phn}^{(+)}$ it is always $\omega_{phn}^{(+)} > \omega_L$, and for the frequency $\omega_{phn}^{(-)}$ we have $\omega_{phn}^{(-)} < \omega_T$. In the most interesting limiting case

$$\omega_n^2 << \omega_T^2 \quad (21)$$

expressions for frequencies (20) are simplified

$$\omega_{phn}^{(-)2} = \frac{\omega_n^2}{d_{st}} \equiv \frac{\omega_n^2 \beta_0^2}{\beta_0^2 \varepsilon_{st} - 1}, \quad (22)$$

$$\omega_{phn}^{(+)2} = \omega_F^2 + \lambda_n^2 \delta\omega_0^2, \quad (23)$$

$$\omega_F^2 = \omega_T^2 \frac{d_{st}}{d_{opt}}, \ \delta\omega_0^2 = \frac{c^2}{b^2} \frac{\Delta\varepsilon}{d_{st} d_{opt}}, \ \Delta\varepsilon = \varepsilon_{st} - \varepsilon_{opt}.$$

In the limit case

$$\omega_n^2 >> \omega_L^2 = \omega_T^2 \frac{\varepsilon_{st}}{\varepsilon_{opt}} \quad (24)$$

expressions for the frequencies of the eigen waves of the dielectric waveguide, synchronous with the electron bunch, follow from (20) and have the form

$$\omega_{phn}^{(-)2} = \omega_T^2 - \omega_n^2 \Delta\varepsilon, \quad (25)$$

$$\omega_{phn}^{(+)2} = \frac{\omega_n^2}{d_{opt}} \equiv \frac{\omega_n^2 \beta_0^2}{\beta_0^2 \varepsilon_{opt} - 1}. \quad (26)$$

Frequency $\omega_{phn}^{(-)}$ (22) is well known in the theory of wakefields excitation by relativistic electron bunch in dielectric waveguides and resonators [1] and is in the microwave (terahertz) range. The frequency $\omega_{phn}^{(+)}$ (23) lies in the infrared range and the process of wakefields excitation at this frequency, as it seems to us, has not been previously studied.

For further analysis, the Fourier integral (15) is conveniently represented as

$$S_n(\bar{t}) = \frac{1}{d_{opt}} \int_{-\infty}^{\infty} \frac{k_\perp^2(\omega)\omega}{k_0^2 \varepsilon(\omega)} \frac{\left(\omega^2 - \omega_T^2\right)e^{-i\omega\bar{t}}}{\left(\omega^2 - \omega_{phn}^{(-)2}\right)\left(\omega^2 - \omega_{phn}^{(+)2}\right)} d\omega.$$

By calculating the residues in the poles $\omega = \pm \omega_{phn}^{(-)} - i0, \ \omega = \pm \omega_{phn}^{(+)} - i0,$ we find

$$S_n(\bar{t}) = -2\pi i \frac{\lambda_n^2}{d_{opt}} \left( \frac{1}{\lambda_n^2 + k_{phn}^{(-)2} b^2} \frac{\omega_T^2 - \omega_{phn}^{(-)2}}{\omega_{phn}^{(+)2} - \omega_{phn}^{(-)2}} \vartheta(\bar{t}) \cos \omega_{phn}^{(-)} \bar{t} + \right.$$

$$\left. + \frac{1}{\lambda_n^2 + k_{phn}^{(+)2} b^2} \frac{\omega_{phn}^{(+)2} - \omega_L^2}{\omega_{phn}^{(+)2} - \omega_{phn}^{(-)2}} \vartheta(\bar{t}) \cos \omega_{phn}^{(+)} \bar{t} \right),$$

where $k_{phn}^{(\mp)} = \omega_{phn}^{(\mp)} / v_0$. Accordingly, for the low-frequency (infrared) part of the electromagnetic Green function, we obtain the following expression

$$E_{Gz}^{(t)}(r, r_0, \bar{t}) = dE_{z-}^{(t)}(r, r_0, \bar{t}) + dE_{z+}^{(t)}(r, r_0, \bar{t}), \quad (27)$$

$$dE_{z-}^{(t)}(r, r_0, \bar{t}) = \frac{dE_w}{d_{opt}} \sum_{n=1}^{\infty} L_n^{(-)} \Pi_n(r, r_0) \vartheta(\bar{t}) \cos \omega_{phn}^{(-)} \bar{t}, \quad (28)$$

$$L_n^{(-)} = \frac{\lambda_n^2}{\lambda_n^2 + k_{phn}^{(-)2} b^2} \frac{\omega_T^2 - \omega_{phn}^{(-)2}}{\omega_{phn}^{(+)2} - \omega_{phn}^{(-)2}},$$

$$dE_{z+}^{(t)}(r, r_0, \overline{t}) = \frac{dE_w}{d_{opt}} \sum_{n=1}^{\infty} L_n^{(+)} \Pi_n(r, r_0) \vartheta(\overline{t}) \cos \omega_{phn}^{(+)} \overline{t}, \quad (29)$$

$$L_n^{(+)} = \frac{\lambda_n^2}{\lambda_n^2 + k_{phn}^{(+)2} b^2} \frac{\omega_{phn}^{(+)2} - \omega_T^2}{\omega_{phn}^{(+)2} - \omega_{phn}^{(-)2}},$$

$$\Pi_n(r, r_0) = \frac{J_0(\lambda_n r / b) J_0(\lambda_n r_0 / b)}{J_1^2(\lambda_n)}, \quad dE_w = \frac{4dQ}{b^2}.$$

In the limiting cases (21), (24) for the coefficients $L_n^{(\mp)}$ the following approximate representations are valid

$$L_n^{(-)} = \begin{cases} \dfrac{d_{opt}}{\varepsilon_{st}}, & \omega_n^2 \ll \omega_T^2, \\ d_{opt} \Delta\varepsilon \dfrac{\omega_T^4}{\omega_n^4}, & \omega_n^2 \gg \omega_L^2, \end{cases}$$

$$L_n^{(+)} = \begin{cases} \beta_0^2 \dfrac{\Delta\varepsilon}{d_{st}} \dfrac{\lambda_n^2}{\kappa_F^2 b^2}, & \omega_n^2 \ll \omega_T^2, \\ \dfrac{d_{opt}}{\varepsilon_{opt}}, & \omega_n^2 \gg \omega_L^2, \end{cases}$$

$\kappa_F = \omega_F / c$. Let us present the expression for the terms of the Green's function (28), (29) in the most interesting case (21)

$$dE_{z-}^{(l)}(r, r_0, \overline{t}) = \frac{dE_w}{\varepsilon_{st}} \sum_{n=1}^{n_{\max}} \Pi_n(r, r_0) \vartheta(\overline{t}) \cos \omega_{phn}^{(-)} \overline{t}, \quad (30)$$

$$dE_{z+}^{(l)}(r, r_0, \overline{t}) = \frac{dE_w \Delta\varepsilon}{d_{st}^2 \kappa_T^2 b^2} \sum_{n=1}^{n_{\max}} \lambda_n^2 \Pi_n(r, r_0) \vartheta(\overline{t}) \cos \omega_{phn}^{(+)} \overline{t}, (31)$$

$\kappa_T = \omega_T / c$.

The upper limit of summation in the sums (30), (31) is determined from the condition $\omega_n \leq \omega_T$ and, taking into account that $\lambda_n \approx \pi(n - 1/4)$, we find

$$n_{\max} \simeq \kappa_T b / \pi \gg 1. \quad (32)$$

Expression (30) describes wake electromagnetic field excited by an infinitely thin electron ring bunch in ion dielectric waveguide in the microwave frequency range, in which there is no frequency dispersion of the dielectric constant. This expression for the wakefield coincides with that obtained in [6]. The exact expression (29) and approximate one (31) describe the excitation of wake electromagnetic waves belonging to a infrared branch of the electromagnetic waves of ion dielectric waveguide. In approximation (21), the frequencies of these waves (23) are practically independent on radial harmonic number.

Thus, we obtained the Green function, which describes the longitudinal component of the wake electric field excited by a ring relativistic electron bunch in ion dielectric waveguide. The Green function contains the longitudinal (potential) and electromagnetic (vortex) parts. In the infrared range, the potential part is a field of longitudinal optical phonons. As for the electromagnetic part of the Green function, it contains a set of radial electromagnetic waves whose frequencies are in the microwave range, as well as electromagnetic radiation in the infrared frequency range.

## 1.2. EXCITATION OF WAKEFIELDS BY AN ELECTRON BUNCH OF FINITE SIZES

The resulting electromagnetic field $\vec{E}(r, \tau)$ of the electron bunch can be found by summing the fields $\vec{E}_G$ of elementary electron rings in accordance with formula (8).

We first consider the excitation of longitudinal optical phonons. Using the potential polarization part of the Green function (18), we obtain the following expression for the wakefield of longitudinal optical phonons

$$E_z^{(l)}(r, \tau) = E_L \Gamma_L(r) Z_\|(\omega_L \tau), \quad (33)$$

where

$$Z_\|(\omega\tau) = \frac{1}{t_{eff}} \int_{-\infty}^{\tau} T(\tau_0 / t_b) \cos \omega(\tau - \tau_0) d\tau_0, \quad (34)$$

$$\Gamma_L(r) = \frac{2\pi}{s_{eff}} \int_0^b R(r_0 / r_b) G(k_L r, k_L r_0) r_0 dr_0, \quad (35)$$

$$E_L = 2Q \frac{k_L^2}{\varepsilon_{eff}}.$$

The function $Z_\|(\omega\tau)$ describes the distribution of the wake field at a frequency $\omega$ in the longitudinal direction at each moment of time. We will consider an electron bunch with a symmetric longitudinal profile $T(\tau_0) = T(-\tau_0)$. The wake function $Z_\|(\omega\tau)$ is conveniently represented as

$$Z_\|(\omega\tau) = \frac{1}{\hat{\tau}} \left[ \hat{T}(\Omega) \vartheta(\tau) \cos \omega\tau - X(\overline{\tau}) \right], \quad (36)$$

where $\Omega = \omega t_b$, $\overline{\tau} = \tau / t_b$,

$$X(\overline{\tau}) = sign\tau \int_{|\overline{\tau}|}^{\infty} T(s) \cos \Omega(|\overline{\tau}| - s) ds, \quad (37)$$

$$\hat{T}(\Omega) = 2 \int_0^{\infty} T(s) \cos(\Omega s) ds, \quad s = t / t_b. \quad (38)$$

The first term (36) describes a wake wave propagating behind the bunch. The amplitude of the wake wave is equal to the Fourier amplitude (38) of the function that describes the longitudinal profile of the electron bunch. The second term in (36) describes the bipolar antisymmetric pulse of the polarization field, localized in the region of the bunch. The field of this pulse tends to zero with distance from the bunch.

Behind a bunch $|\overline{\tau}| \gg 1$, the wake field (33) of longitudinal optical phonons has the form of a monochromatic wave

$$E_z^{(l)}(r, \tau) = E_L \Gamma_L(r) \frac{\hat{T}(\Omega_L)}{\hat{\tau}} \cos \omega_L \tau, \quad \Omega_L = \omega_L t_L. \quad (39)$$

We present the expressions for the Fourier amplitudes $\hat{T}(\Omega_L)$ for two model longitudinal profiles of the electron bunch: Gaussian and power laws

$$T(\tau_0/t_b) = e^{-\tau_0^2/t_b^2}, \quad \hat{T}(\Omega) = \sqrt{\pi}e^{-\Omega^2/4}, \quad \hat{\tau} = \sqrt{\pi}, \quad (40)$$

$$T(\tau_0/t_b) = \frac{1}{1+\tau_0^2/t_b^2}, \quad \hat{T}(\Omega) = \pi e^{-\Omega}, \quad \hat{\tau} = \pi.$$

Longitudinal optical phonons are most efficiently radiated when the coherence condition is fulfilled $\omega_L t_b \leq 1$. If the inequality $\omega_L t_b \gg 1$ takes place, then the longitudinal optical phonons are not coherently radiated and the amplitude of the wake wave is exponentially small.

Let's consider an electron bunch with a Gaussian transverse profile

$$R(r_0/r_b) = e^{-r_0^2/r_b^2}. \quad (41)$$

When the condition $k_L b \gg 1$ is satisfied on the axis $r = 0$ the function $\Gamma_L(r)$ takes on the value

$$\Gamma_L(0) = -\frac{1}{2}e^{\rho_b} Ei(-\rho_b), \rho_b = \frac{k_L^2 r_b^2}{4}, \quad (42)$$

$$Ei(z) = \int_{-\infty}^{z} \frac{e^t}{t} dt$$

is integral exponential function. For thin $\rho_b \ll 1$ and wide $\rho_b \gg 1$ bunches the asymptotic representations for function (37) are

$$\Gamma_L(0) = \begin{cases} \frac{1}{2}\ln\left(\frac{1}{\rho_b}\right), & \rho_b \ll 1, \\ \frac{1}{\rho_b}, & \rho_b \gg 1. \end{cases}$$

Thus, with the full coherence of the Cherenkov radiation of longitudinal optical phonons $\omega_L t_b \leq 1$, $k_L r_b \leq 1$ the wakefield of optical phonons on the axis of the waveguide takes the maximum value

$$E_z^{(l)}(r,\tau) = E_L \ln(2/k_L r_b) \cos \omega_L \tau. \quad (43)$$

We present the expressions for the amplitude of the wakefield on the axis of the waveguide $E_L$ for two ion dielectrics of the alkali-halogen group: sodium chloride NaCl and potassium iodide KI. For sodium chloride, we have

$$E_L = 0.2 N_0 (V/cm),$$

Here $N_0 = Q/e$ is the number of electrons in the bunch. The frequency of longitudinal optical phonons is equal $f_L = 7.62 \cdot 10^{12} Hz$. For $N_0 = 10^9$ from this formula we obtain the estimation for the electric field strength $E_L = 0.2 GV/cm$. Accordingly, for potassium iodide we have

$$E_L = 0.11 N_0 (V/\quad).$$

The frequency of longitudinal optical phonons and the electric field strength are equal $f_L = 4 \cdot 10^{12} Hz$, $E_L = 0.11 GV/cm$, lower than in the previous case.

Let us now consider the excitation of wake electromagnetic waves by an electron bunch. Using the electromagnetic Green function, we obtain the wake electromagnetic field as a superposition of radial modes

$$E_{z(\mp)}^{(t)}(r,\tau) = \frac{E_w}{d_{opt}} \sum_{n=1}^{\infty} L_n^{(\mp)} \Gamma_n \frac{J_0\left(\lambda_n \frac{r}{b}\right)}{J_1^2(\lambda_n)} Z_n^{(\mp)}(\tau), \quad (44)$$

$$\Gamma_n = \frac{2\pi}{s_{eff}} \int_0^b R\left(\frac{r_0}{r_b}\right) J_0\left(\lambda_n \frac{r}{b}\right) r_0 dr_0, \quad Z_n^{(\mp)}(\tau) = Z_{\parallel}(\omega_{phn}^{(\mp)}\tau),$$

$$E_w = 4Q/b^2.$$

For a symmetric electron bunch in the "wave zone" $\omega_{phn}^{(\mp)} \tau \gg 1$, where the quasi-static field of the electron bunch is small, the wake field is a superposition of radial modes of the dielectric waveguide

$$E_z^{(t)}(r,\tau) = E_{z-}^{(t)}(r,\tau) + E_{z+}^{(t)}(r,\tau),$$

$$E_{z(\mp)}^{(t)}(r,\tau) = \frac{E_w}{d_{opt}} \sum_{n=1}^{\infty} L_n^{(\mp)} \Gamma_n \frac{\hat{T}_n^{(\mp)}}{\hat{\tau}} \frac{J_0\left(\lambda_n \frac{r}{b}\right)}{J_1^2(\lambda_n)} \cos \omega_{phn}^{(\mp)} \tau, (45)$$

where $\hat{T}_n^{(\mp)} \equiv \hat{T}(\Omega_{phn}^{(\mp)})$ is Fourier component (38) of the function $T(s)$ at the dimensionless frequency $\Omega_{phn}^{(\mp)} = \omega_{phn}^{(\mp)} t_b$.

We also present an expression for the power of the wake electromagnetic radiation, which we define as the component of the full Poiting vector along the axis of the dielectric waveguide

$$P_{tz} = \frac{c}{4\pi} \int_0^b \langle E_r^{(t)} H_\varphi^{(t)} \rangle 2\pi r dr.$$

Angle brackets mean average over high-frequency wakefield oscillations. The Poiting vector contains two terms

$$P_{tz} = P_{tz}^{(-)} + P_{tz}^{(+)},$$

corresponding to two branches of the polariton oscillations

$$P_{tz}^{(\mp)} = \frac{c}{8} \beta_0 \frac{E_w^2 b^2}{d_{opt}^2} \sum_{n=1}^{\infty} \frac{\omega_n^2 b^2}{v_0^2} \varepsilon_n^{(\mp)} L_n^{(\pm)^2} \Gamma_n^2 \frac{\hat{T}_n^{(\mp)^2}}{\hat{\tau}^2} \frac{J_1^2(\lambda_n)}{\lambda_n^2}, \quad (46)$$

$$\varepsilon_n^{(\mp)} \equiv \varepsilon(\omega_{phn}^{(\mp)}) = \beta_0^{-2} + \frac{\omega_n^2}{\omega_{phn}^{(\mp)^2}}.$$

For an electron bunch with a Gaussian longitudinal (40) and transverse (41) profiles, the coefficients $\hat{T}_n^{(\mp)}$ and $\Gamma_n$, which are determined by the specific form of the bunch density profiles, have the form

$$\hat{T}_n^{(\mp)} = \sqrt{\pi} \exp\left(-\frac{1}{4}\Omega_{phn}^{(\mp)^2}\right), \quad (47)$$

$$\Gamma_n = 2\int_0^{1/\eta_b} J_0(\lambda_n \eta_b \rho) e^{-\rho^2} \rho d\rho, \quad \eta_b = r_b/b.$$

When the condition $\eta_b \ll 1$ is satisfied, the expression for the coefficient $\Gamma_n$ has form

$$\Gamma_n = \exp\left(-\frac{1}{4}\lambda_n^2 \eta_b^2\right). \qquad (48)$$

Accordingly, expression (45) for a Gaussian bunch, taking into account relations (47), (48), takes the form

$$E_{z(\mp)}^{(t)} = \frac{E_w}{d_{opt}} \sum_{n=1}^{\infty} L_n^{(\mp)} e^{-\frac{1}{4}\left(\Omega_{phn}^{(\mp)^2} + \lambda_n^2 \eta_b^2\right)} \frac{J_0\left(\lambda_n \frac{r}{b}\right)}{J_1^2(\lambda_n)} \cos \omega_{phn}^{(\mp)} \tau. \quad (49)$$

For the radiated power, we have

$$P_{tz}^{(\mp)} = \frac{c}{8}\beta_0 \frac{E_w^2 b^2}{d_{opt}^2} \sum_{n=1}^{\infty} \frac{\omega_n^2 b^2}{v_0^2} e^{-\frac{1}{2}\left(\Omega_{phn}^{(\mp)^2} + \lambda_n^2 \eta_b^2\right)} \frac{\varepsilon_n^{(\mp)} L_n^{(\pm)^2}}{\lambda_n^2 J_1^2(\lambda_n)}.$$

From expression (49) it follows that an electron bunch excites finite number of radial modes of electromagnetic radiation, for which the coherence condition $\omega_{phn}^{(\mp)^2} t_b^2 \leq 1$, $\lambda_n^2 r_b^2 / b^2 \leq 1$ for excitation by an electron bunch is satisfied. If the condition for the duration of the electron bunch is satisfied $b/c \gg t_b \gg 1/\omega_L$, then the amplitudes of all radial harmonics of the high-frequency field $E_{z+}^{(t)}$ are exponentially small. Therefore, the electron bunch under these conditions will excite only low-frequency transverse polaritons. Let us consider the most interesting case of an electron bunch of short duration

$$ct_b / b \ll \omega_L t_b \ll 1, \qquad (50)$$

when transverse polaritons belonging to both the low-frequency and high-frequency branches of oscillations will be coherently excited.

In the limiting case (21) for the frequencies of the radial harmonics, we can use the approximate expressions (22), (23). As a result, for wakefields excited by electron bunches with Gaussian longitudinal and transverse profiles, instead of exact expressions (49), we obtain approximate relations

$$E_{z-}^{(t)}(r,\tau) = \frac{E_w}{\varepsilon_{st}} \sum_{n=1}^{n_{max}} e^{-\frac{1}{4}\left(\lambda_n^2 \eta_b^2 + \Omega_{phn}^{(-)^2}\right)} \frac{J_0\left(\lambda_n \frac{r}{b}\right)}{J_1^2(\lambda_n)} \cos \omega_{phn}^{(-)} \tau, (51)$$

$$E_{z+}^{(t)}(r,\tau) = E_w \frac{\beta_0^2 \Delta\varepsilon}{d_{opt} d_{st}} e^{-\frac{\Omega_F^2}{4}} \frac{1}{\kappa_F^2 b^2} \mathrm{Re}\left[e^{-i\omega_F \bar{\tau}} \Lambda(r,\tau)\right], (52)$$

where $\Omega_F = \omega_F t_b$. The function $\Lambda(r,\bar{t})$ is defined as follows

$$\Lambda(r,\tau) = \sum_{n=1}^{n_{max}} \frac{\lambda_n^2}{J_1^2(\lambda_n)} e^{-\frac{\lambda_n^2}{4}\left(\eta_b^2 + \delta\omega_0^2 t_b^2\right) - i\lambda_n^2 \Delta\omega_0 \tau} J_0\left(\lambda_n \frac{r}{b}\right). \quad (53)$$

Here $\Delta\omega_0 = \delta\omega_0^2 / 2\omega_F$. In the sums (51) (53), we took into account only the terms with numbers $n < n_{max}$, where $n_{max}$ is defined by formula (32).

## 2. EXCITATION OF SURFACE POLARITONS BY A RELATIVISTIC ELECTRONIC BUNCH IN A DIELECTRIC WAVEGUIDE

The waveguide system is performed in the form of a dielectric tube. The inner region $r < a$ is a vacuum cylindrical cavity. The region $b > r > a$ is filled with a homogeneous ion dielectric. The outer surface of the dielectric tube is covered with a perfectly conductive metal film that completely shields the electromagnetic field in the dielectric waveguide. A relativistic electron bunch propagates in the vacuum channel

### 2.1. DISPERSION PROPERTIES OF SURFACE WAVES OF A DIELECTRIC WAVEGUIDE

Let us dwell briefly on the dispersion properties of surface waves in a dielectric waveguide. The surface waves of the considered dielectric waveguide are described by the equation

$$\varepsilon(\omega) = -\frac{1}{q_v a} \frac{I_1(q_v a)}{I_0(q_v a)} q_d a \frac{\Delta_0(q_d a, q_d b)}{\Delta_1(q_d a, q_d b)}, \qquad (54)$$

in which

$$q_v = \sqrt{k^2 - k_0^2}, \quad q_d = \sqrt{k^2 - k_0^2 \varepsilon(\omega)}, \quad k_0 = \omega/c,$$

$$\Delta_n(q_d r, q_d b) = I_0(q_d b) K_n(q_d r) - (-1)^n I_n(q_d r) K_0(q_d b),$$

where $k$ is the longitudinal wave number. It is convenient to classify waves propagating in a dielectric waveguide using the diagram on the plane $\omega, k$, which is shown in Fig. 1 [22-24].

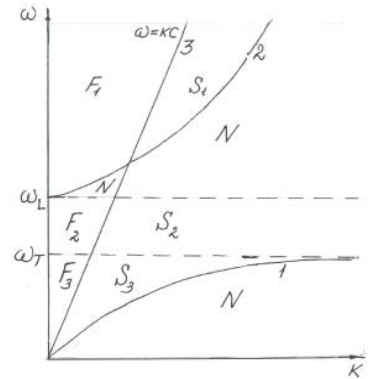

Fig.1. Regions of existence of various types waves in an ion dielectric waveguide

Boundary curves 1, 2 are the dispersion dependences of volume polaritons in an infinite ion dielectric, the "light line" $\omega = kc$ divides the plane $\omega, k$ into regions of existence of fast $v_{ph} = \omega/k > c$ (regions $F$) and slow $v_{ph} < c$ (regions $S$) eigen waves of the dielectric waveguide. To the left of the light line in the regions $F_{1,3}$ there are fast volume $q_{d,v}^2 < 0$ eigen waves of the dielectric waveguide. The frequency range of fast volume waves in the region $F_1$ is determined from the condition $\varepsilon(\omega) < 1$ from which it follows

$$\omega_L \sqrt{\frac{\varepsilon_{opt}(\varepsilon_{st}-1)}{\varepsilon_{st}(\varepsilon_{opt}-1)}} > \omega > \omega_L.$$

The region $F_2$ corresponds to fast waves, that propagate in a vacuum channel and experience full internal reflection from the vacuum-dielectric boundary. In a vacuum, these waves are volume $q_v^2 < 0$, and in a dielectric they are surface waves $q_d^2 > 0$. This region is limited by inequalities $\omega > kc$, $\omega_L > \omega > \omega_T$.

The regions $S_{1,3}$ located to the right of the light line correspond to slow waves, which are volume $q_d^2 < 0$ in the dielectric and surface $q_v^2 > 0$ in the vacuum channel. And finally, the region $S_2$ corresponds to slow purely surface waves $q_{d,v}^2 > 0$. This region is defined by the inequalities

$$\omega < kc, \quad \omega_L > \omega > \omega_T. \qquad (55)$$

The dispersion equation (54) can be significantly simplified in the limit case of a large radius of the vacuum channel $q_{v,d} a \gg 1$. This limit case corresponds to the transition from an annular waveguide to a flat dielectric layer. The dispersion equation for a flat dielectric layer takes the form

$$\varepsilon(\omega) = -\frac{q_d}{q_v}\tanh(q_d L), \qquad (56)$$

where $L = b - a$ is the layer thickness. The results of the analysis of this equation with the dependence of the dielectric constant on frequency (17) are presented, for example, in works [19, 24]. In turn, in the case of a thick layer $q_d \gg 1$, dispersion equation (56) takes the form

$$\varepsilon(\omega) = -\frac{q_d}{q_v} \qquad (57)$$

and describes the surface waves of the dielectric half-space. When $k \to \infty$ dispersion equations (56), (57) go over to the equation

$$\varepsilon = -1,$$

which determines the limit frequency

$$\omega_\infty = \omega_T \sqrt{\frac{\varepsilon_{st}+1}{\varepsilon_{opt}+1}}, \quad \omega_L > \omega_\infty > \omega_T. \qquad (58)$$

For every thickness of the dielectric layer at $k \to \infty$, the dispersion curve tends to the limit frequency $\omega_\infty$.

Let us now consider the behavior of the dispersion curve $\omega(k)$ in the vicinity of the left boundary $\omega = ck$ of the region of existence of surface waves (see Fig. 1). In this vicinity, the phase velocities of surface waves are close to the speed of light in vacuum; therefore, these waves are of the greatest interest from the point of view of their wake excitation by relativistic electron bunches. In this limiting case, dispersion equation (56) takes the form

$$\varepsilon(\omega) = -\frac{\kappa_0 \sqrt{1-\varepsilon(\omega)}}{q_v}\tanh\left(\kappa_0 L\sqrt{1-\varepsilon(\omega)}\right). \qquad (59)$$

It follows from this equation that at $q_v \to 0$ frequency $\omega \to \omega_T$, i.e. for every thickness of the dielectric layer, the dispersion curve on the straight line $\omega = ck$ begins from the point with coordinates $\omega = \omega_T$, $k_T = \omega_T/c$. Numerical analysis of dispersion equation (56) shows [19, 24], that for thin dielectric layers the dispersion curves have a segment with anomalous dispersion. To obtain a criterion for the appearance on the dispersion curve of a region with anomalous dispersion, we use the following asymptotic representation of the dispersion dependence $\omega(k)$ of a surface wave at $k \to \infty$

$$\omega(x) = \omega_\infty\left[1 - \frac{\Delta\varepsilon}{x^2}f(x)\right], \qquad (60)$$

where $x = 2kL$,

$$\Delta\varepsilon = \frac{\varepsilon_{st}-\varepsilon_{opt}}{(\varepsilon_{st}+1)(\varepsilon_{opt}+1)}, \quad f(x) = x_L - x^2 e^{-x},$$

$$x_L = \frac{2\omega_\infty^2 L^2}{c^2}.$$

It follows from relation (60) that if for all values $x \gg 1$ the function $f(x)$ is positive definite

$$f(x) > 0, \qquad (61)$$

then the dispersion curve always approaches to the limiting frequency $\omega_\infty$ from below. In this case, normal dispersion of the surface wave takes place. Condition (61) is always satisfied for the dielectric layer with parameters

$$\frac{\omega_\infty L}{c} > \frac{\sqrt{2}}{e} \approx 0.52. \qquad (62)$$

If the layer is thin and the inequality, which is opposite to (62), is satisfied, then the function $f(x)$ becomes alternating and a segment with anomalous dispersion appears on the dispersion curve $\omega(k)$.

Let us now consider the features of the dispersion of surface waves in a tubular dielectric waveguide. At $k \to \infty$, the dispersion curve, as in the case of a flat layer, always reaches the limit frequency (58). As for the behavior of the dispersion curve near the left boundary of the region of existence of a surface wave $\omega = kc$, it differs radically from the case of a plane layer. In the vicinity of this boundary, the condition

$$q_v a \ll 1 \qquad (63)$$

is satisfied. This condition makes it possible to significantly simplify dispersion equation (55) and represent it in the form

$$\varepsilon(\omega) = -\frac{q_d a}{2}\frac{\Delta_0(q_d a, q_d b)}{\Delta_1(q_d a, q_d b)}. \qquad (64)$$

Since the right-hand side of this equation as a function $q_d$ can change from 0 to $-\infty$ the value of the frequency $\omega_b$ at which the dispersion curve of surface polaritons begins changes over the all range of existence of surface polaritons $\omega_L > \omega_b > \omega_T$.

If, along with condition (63), the condition is satisfied too
$$q_d a \gg 1, \quad (65)$$
then instead of equation (59) in the vicinity of the boundary $\omega = kc$ we obtain
$$\varepsilon(\omega) = -\frac{k_0 a \sqrt{1-\varepsilon(\omega)}}{2} \tanh\left(k_0 L \sqrt{1-\varepsilon(\omega)}\right). \quad (66)$$
This equation differs significantly from the analogous equation (59) for a plane layer. In the limiting case
$$k_0 L \sqrt{1-\varepsilon(\omega)} \gg 1$$
of a thick layer, for determination the initial frequency $\omega_b$ equation (66) is simplified and takes the form
$$\frac{\varepsilon(\omega)}{\sqrt{1-\varepsilon(\omega)}} = -\frac{k_0 a}{2}.$$
Since the right side of this equation is large, from this equation we find the approximate value of the frequency at which the dispersion curve starts at frequency
$$\omega = \omega_b \equiv \omega_T \left(1 + \frac{2\Delta\varepsilon}{\kappa_T^2 a^2}\right). \quad (67)$$
The condition for the applicability of this formula takes the form
$$\kappa_T^2 a^2 / 2 \gg 1.$$
Thus, in the case of a thick layer, the dispersion curve of the surface wave on the boundary straight line $\omega = ck$ begins from the frequency (67), which slightly exceeds the frequency of transverse phonons.

Let's consider now the limiting case of a thin dielectric layer
$$k_0 L \sqrt{1-\varepsilon(\omega)} \ll 1.$$
In this case, the equation (66) for determination the initial frequency $\omega_b$ is simplified and takes the form
$$\frac{\varepsilon(\omega)}{1-\varepsilon(\omega)} = -\frac{k_0^2 aL}{2}. \quad (68)$$
The right side of this equation is the product of the large $k_0 a \gg 1$ and small $k_0 L \ll 1$ parameters. Therefore, it can take, generally speaking, an arbitrary value. Equation (68) is equivalent to the following
$$\frac{\omega^2 - \omega_L^2}{\omega^2 - \omega_{F0}^2} = \omega^2 \frac{aL}{2c^2} \frac{\varepsilon_{opt} - 1}{\varepsilon_{opt}},$$
where
$$\omega_{F0}^2 = \omega_T^2 \frac{\varepsilon_{st} - 1}{\varepsilon_{opt} - 1} < \omega_L^2.$$
The root of this equation has form
$$\omega_b^2 = \frac{2\omega_L^2}{1 + \alpha_{T0} + \sqrt{(1+\alpha_{T0})^2 - 4\alpha_{T0}\epsilon_0}}.$$
Here
$$\alpha_{T0} = \frac{\kappa_T^2 aL}{2} \frac{\varepsilon_{st} - 1}{\varepsilon_{opt}}, \quad (69)$$
$$\epsilon_0 = \frac{\omega_L^2}{\omega_{F0}^2} = \frac{\varepsilon_{st}}{\varepsilon_{opt}} \frac{\varepsilon_{opt} - 1}{\varepsilon_{st} - 1} < 1.$$
This initial frequency belongs to the region of existence of surface waves $\omega_L > \omega > \omega_T$ for the following values of the parameters of the dielectric layer
$$\frac{\kappa_T^2 aL}{2} < 1.$$
If the dielectric layer is so thin that the condition $\kappa_T^2 aL / 2 \ll 1$ is satisfied, then the initial frequency is close to the frequency of longitudinal optical phonons
$$\omega_b^2 = \omega_L^2 \left(1 - \frac{\kappa_T aL}{2} \frac{\Delta\varepsilon}{\varepsilon_{opt}^2}\right).$$

Finally, we consider the limiting case of vacuum channel of the a small radius
$$q_d a \ll 1, \quad q_d b \geq 1.$$
In this case, from dispersion equation (54) we obtain the following equation for determination the initial frequency
$$\frac{\varepsilon(\omega)}{1-\varepsilon(\omega)} = -\frac{k_0^2 a^2}{2} \ln\left(\frac{1}{\nu k_0 a [1-\varepsilon(\omega)]}\right),$$
$\nu$ is Euler's constant. Solving this equation by the method of successive approximations, we find the value of the initial frequency
$$\omega_b = \omega_L - \frac{1}{4\varepsilon_{eff}} \kappa_L^2 a^2 \ln\frac{1}{\nu\kappa_L a},$$
where $\kappa_L = \omega_L / c$. The initial frequency at which the dispersion curve begins is slightly lower than the frequency of longitudinal optical phonons. In this case, the longitudinal wave number is $\kappa_b = \omega_b / c$. If a more stringent condition is satisfied, namely
$$q_d a \ll 1, \quad q_d b \ll 1,$$
then we have equation
$$\frac{\varepsilon(\omega)}{1-\varepsilon(\omega)} = -\frac{k_0^2 a^2}{2} \ln\left(\frac{b}{a}\right).$$
Accordingly, for the initial frequency, we obtain
$$\omega_b = \omega_L - \frac{1}{4\varepsilon_{eff}} \kappa_L^2 a^2 \ln\frac{b}{a}.$$

Thus, for the considered tubular dielectric waveguide, depending on the thickness of the dielectric layer, the initial frequency of the dispersion curve of surface polaritons can take any value in the frequency range of their existence $\omega_L > \omega > \omega_T$.

## 2.2. DETERMINATION OF THE GREEN FUNCTION IN A TUBULAR DIELECTRIC WAVEGUIDE

We will solve the problem by the same method as in the previous section. Let us first determine the wake field of an electron bunch in the form of an infinitely thin ring (2). The longitudinal component of the Fourier amplitude of the field $E_{Gz\omega}^{(v)}$ in the vacuum channel $r < a$ is described by the equation

$$\frac{1}{r}\frac{d}{dr}r\frac{dE_{Gz\omega}^{(v)}}{dr} - \kappa_v^2 E_{Gz\omega}^{(v)} = -\frac{i}{\pi}\frac{\omega}{v_0^2\gamma_0^2}dQ\frac{\delta(r-r_0)}{r_0}, \quad (70a)$$

In the dielectric electron bunch is absent. Therefore, the equation for the Fourier amplitude is homogeneous

$$\frac{1}{r}\frac{d}{dr}r\frac{dE_{Gz\omega}^{(d)}}{dr} - \kappa_d^2 E_{Gz\omega}^{(d)} = 0, \quad (70b)$$

where

$$\kappa_v^2 = k_l^2/\gamma_0^2, \; \kappa_d^2 = k_l^2 - k_0^2\varepsilon \equiv k_l^2\left(1-\beta_0^2\varepsilon\right), \; k_l = \omega/v_0.$$

Let's split the dielectric waveguide into three regions: two regions in the vacuum channel

$$E_{Gz\omega}^{(v1)}(r) = E_{Gz\omega}^{(v)}(r < r_0 < a), \; E_{Gz\omega}^{(v2)}(r) = E_{Gz\omega}^{(v)}(a > r > r_0)$$

and the third region is a dielectric tube $b > r > a$. The electromagnetic field must satisfy the following boundary conditions. On the surface of a perfectly conducting side metal film, the longitudinal component of the electric field is equal zero

$$E_{Gz\omega}^{(d)}(r=b) = 0.$$

The tangential components of the electric and magnetic fields are continuous at the dielectric-vacuum boundary

$$E_{Gz\omega}^{(d)}(r=a) = E_{Gz\omega}^{(v2)}(r=a), \quad (71)$$

$$H_{G\varphi\omega}^{(d)}(r=a) = H_{G\varphi\omega}^{(v2)}(r=a). \quad (72)$$

In the considered case of an infinitely thin ring bunch, the Fourier amplitude of the magnetic field $H_{G\varphi\omega}^{(v)}(r)$ undergoes a jump on the surface $r = r_0$ along which the ring bunch propagates

$$H_{G\varphi\omega}^{(v2)}(r=r_0) - H_{G\varphi\omega}^{(v1)}(r=r_0) = -\frac{dQ}{\pi c r_0}. \quad (73)$$

The longitudinal component of the electric field on this surface is continuous

$$E_{Gz\omega}^{(v2)}(r=r_0) = E_{Gz\omega}^{(v1)}(r=r_0). \quad (74)$$

The solution of equations (70) in each of the regions of the dielectric waveguide, as well as the corresponding Fourier components of the magnetic field, have the form

$$E_{Gz\omega}^{(v1)}(r) = A_1 I_0(\kappa_v r), \; H_{G\varphi\omega}^{(v1)}(r) = -i\frac{k_0}{\kappa_v}A_1 I_1(\kappa_v r),$$

$$E_{Gz\omega}^{(v2)}(r) = A_2 I_0(\kappa_v r) + A_3 K_0(\kappa_v r),$$

$$H_{G\varphi\omega}^{(v2)}(r) = -i\frac{k_0}{\kappa_v}\left[A_2 I_1(\kappa_v r) - A_3 K_1(\kappa_v r)\right],$$

$$E_{Gz\omega}^{(d)}(r) = A_4 \Delta_0(\kappa_d r, \kappa_d b),$$

$$H_{G\varphi\omega}^{(d)}(r) = -i\frac{k_0\varepsilon(\omega)}{\kappa_d}A_4 \Delta_1(\kappa_d r, \kappa_d b),$$

where

$$\Delta_n(\kappa_d r, \kappa_d b) = I_0(\kappa_d b)K_n(\kappa_d r) - (-1)^n I_n(\kappa_d r)K_0(\kappa_d b).$$

The coefficients $A_n$, $n = 1 \div 4$ are to be determined from the boundary conditions (71) - (74). Using these conditions, we obtain the system of inhomogeneous linear equations for determination of the coefficients $A_n$. Having obtained a solution to this system and performing the inverse Fourier transform, we find expressions for the components of the electromagnetic field in each of the regions of the dielectric waveguide in the form of integral Fourier representations

$$E_{Gz}^{(v1)} = \frac{i}{\pi}\frac{dQ}{a^2}\int_{-\infty}^{\infty}e^{-i\omega\tilde{t}}\frac{d\omega}{\omega}\frac{I_0(\kappa_v r)}{I_0(\kappa_v a)}\left[\frac{I(\kappa_v r_0)}{I_0(\kappa_v a)}\frac{1}{D(\omega)}+\right.$$
$$\left.+\kappa_v^2 a^2 \Delta_0(\kappa_v r_0, \kappa_v a)\right], \quad (75)$$

$$H_{G\varphi}^{(v1)} = \frac{1}{\pi}\frac{dQ}{a^2}\int_{-\infty}^{\infty}e^{-i\omega\tilde{t}}\frac{d\omega}{\kappa_v c}\frac{I_1(\kappa_v r)}{I_0(\kappa_v a)}\left[\frac{I_0(\kappa_v r_0)}{I_0(\kappa_v a)}\frac{1}{D(\omega)}+\right.$$
$$\left.+\kappa_v^2 a^2 \Delta_0(\kappa_v r_0, \kappa_v a)\right],$$

$$E_{Gz}^{(v2)}(r) = \frac{i}{\pi}\frac{dQ}{a^2}\int_{-\infty}^{\infty}e^{-i\omega\tilde{t}}\frac{d\omega}{\omega}\frac{I_0(\kappa_v r_0)}{I_0(\kappa_v a)}\left[\frac{I_0(\kappa_v r)}{I_0(\kappa_v a)}\frac{1}{D(\omega)}+\right.$$
$$\left.+\kappa_v a \Delta_0(\kappa_v r, \kappa_v a)\right], \quad (76)$$

$$H_{G\varphi}^{(v2)}(r) = \frac{1}{\pi}\frac{dQ}{a^2}\int_{-\infty}^{\infty}e^{-i\omega\tilde{t}}\frac{d\omega}{\kappa_v c}\frac{I_0(\kappa_v r_0)}{I_0(\kappa_v a)}\left[\frac{I_1(\kappa_v r)}{I_0(\kappa_v a)}\frac{1}{D(\omega)}-\right.$$
$$\left.-\kappa_v^2 a^2 \Delta_1(\kappa_v r, \kappa_v a)\right],$$

$$E_{Gz}^{(d)} = \frac{i}{\pi}\frac{dQ}{a^2}\int_{-\infty}^{\infty}\frac{I_0(\kappa_v r_0)}{I_0(\kappa_v a)}\frac{\Delta_0(\kappa_d r, \kappa_d b)}{\Delta_0(\kappa_d a, \kappa_d b)}\frac{e^{-i\omega\tilde{t}}d\omega}{\omega D(\omega)}, \quad (77)$$

$$H_{G\varphi}^{(d)} = -\frac{1}{\pi}\frac{dQ}{a^2}\int_{-\infty}^{\infty}\frac{I_0(\kappa_v r_0)}{I_0(\kappa_v a)}\frac{\Delta_1(\kappa_d r, \kappa_d b)}{\Delta_0(\kappa_d a, \kappa_d b)}\frac{\varepsilon(\omega)e^{-i\omega\tilde{t}}d\omega}{\kappa_d\,(\omega)},$$

where

$$D(\omega) = \frac{1}{\kappa_v a}\frac{I_1(\kappa_v a)}{I_0(\kappa_v a)} + \frac{\varepsilon}{\kappa_d a}\frac{\Delta_1(\kappa_d a, \kappa_d b)}{\Delta_0(\kappa_d a, \kappa_d b)}. \quad (78)$$

The integrands in (75) - (77) have only simple poles $\omega = \pm\omega_s - i0$, which are the roots of the equation

$$D(\omega) = 0. \quad (79)$$

Note that the integral Fourier representations (75) - (77) for the wakefield are valid for all isotropic media (dielectrics, gas plasma, plasma of semiconductors and metals), the polarization properties of which are described by the scalar dielectric constant $\varepsilon(\omega)$. The roots of equation (79) determine the frequencies of the eigin waves of the dielectric waveguide, synchronous with the relativistic electron bunch. In the framework of the considered model, a dielectric waveguide has a formally infinite number of branches of slow volume electromagnetic waves (volume polaritons) and one surface electromagnetic wave (surface polariton). Cherenkov excitation by a relativistic electron bunch of volume polaritons was considered in the previous section for the model of a dielectric waveguide with full filling. The presence of a vacuum channel does not introduce qualitative features into the picture of the excitation of volume polaritons; therefore, below we will focus on investigation of the physical picture of only surface polaritons excitation. Note that there is no pole $\varepsilon(\omega) = 0$ in the integrands (75) - (77). Therefore, an electron bunch moving in a vacuum channel does not excite longitudinal optical phonons. A similar situation takes place with wake excitation of surface waves in a plasma waveguide [25].

Calculating the residues in the poles $\omega = \pm\omega_s - i0$, we find the expression for the longitudinal component of the electric field of the surface wave

$$E_{Gz}^{(v,s)} = \frac{2dQ}{a^2}\frac{1}{\Lambda_s}\frac{I_0(\kappa_{vs}r_0)I_0(\kappa_{vs}r)}{I_0^2(\kappa_{vs}a)}\vartheta(\bar{t})\cos\omega_s\bar{t}. \quad (80)$$

where $\kappa_{vs} \equiv \kappa_v(\omega_s)$, $\Lambda_s \equiv \Lambda(\omega_s)$,

$$\Lambda(\omega) = \omega^2 \frac{\partial D(\omega^2)}{\partial \omega^2}. \quad (81)$$

Accordingly, the expression for the magnetic field of the surface wave has the form

$$H_{G\varphi}^{(v,s)} = -\frac{2dQ}{a^2}\beta_0\gamma_0\frac{1}{\Lambda_s}\frac{I_0(\kappa_{vs}r_0)I_1(\kappa_{vs}r)}{I_0^2(\kappa_{vs}a)}\vartheta(\bar{t})\sin\omega_s\bar{t}.$$

We also give an expression for these components of the electromagnetic field in the dielectric

$$E_{Gz}^{(d,s)} = \frac{2dQ}{a^2}\frac{1}{\Lambda_s}\frac{I_0(\kappa_{vs}r_0)}{I_0(\kappa_{vs}a)}\times$$

$$\times\frac{\Delta_0(\kappa_{ds}r,\kappa_{ds}b)}{\Delta_0(\kappa_{ds}a,\kappa_{ds}b)}\vartheta(\bar{t})\cos\omega_s\bar{t},)$$

$$H_{G\varphi}^{(d,s)} = \frac{2dQ}{a^2}\frac{1}{\Lambda_s}\frac{I_0(\kappa_{vs}r_0)}{I_0(\kappa_{vs}a)}\frac{\omega_s}{\kappa_{ds}c}\frac{\Delta_1(\kappa_{ds}r,\kappa_{ds}b)}{\Delta_0(\kappa_{ds}a,\kappa_{ds}b)}\vartheta(\bar{t})\sin\omega_s\bar{t},$$

where $\kappa_{ds} = \kappa_d(\omega_s)$.

Thus, we have obtained expressions that describe a wake surface wave excited by a ring electron bunch in an isotropic dielectric waveguide with an axial vacuum channel.

## 2.2. EXCITATION OF SURFACE WAKEFIELDS BY AN ELECTRON BUNCH OF FINITE SIZES

The field of a surface wave excited by an electron bunch of finite sizes is found by summing of the fields of elementary ring charges and currents according to formula (8). As a result, we obtain the following expressions for the components of the electromagnetic field of the surface wave

$$E_z^{(v,s)} = \frac{2Q}{a^2}\frac{1}{\Lambda_s}\frac{I_0(\kappa_{vs}r)}{I_0(\kappa_{vs}a)}\Gamma_{s0}Z_{\parallel}(\omega_s\tau), \quad (82)$$

$$H_{G\varphi}^{(v,s)} = -\frac{2Q}{a^2}\beta_0\gamma_0\frac{1}{\Lambda_s}\frac{I_1(\kappa_{vs}r)}{I_0(\kappa_{vs}a)}\Gamma_{s0}Z_{\perp}(\omega_s\tau),$$

$$E_z^{(d,s)} = \frac{2Q}{a^2}\frac{1}{\Lambda_s}\frac{\Delta_0(\kappa_{ds}r,\kappa_{ds}b)}{\Delta_0(\kappa_{ds}a,\kappa_{ds}b)}\Gamma_{s0}Z_{\parallel}(\omega_s\tau),$$

$$H_{\varphi}^{(d,s)} = \frac{2Q}{a^2}\frac{1}{\Lambda_s}\frac{\varepsilon(\omega_s)\omega_s}{\kappa_{ds}c}\frac{\Delta_1(\kappa_{ds}r,\kappa_{ds}b)}{\Delta_0(\kappa_{ds}a,\kappa_{ds}b)}\Gamma_{s0}Z_{\perp}(\omega_s\tau),$$

where

$$Z_{\perp}(\omega_s\tau) = \frac{1}{t_{eff}}\int_{-\infty}^{\tau}T(\tau_0/t_b)\sin\omega_s(\tau-\tau_0)d\tau_0,$$

$$\Gamma_{s0} = \frac{2\pi}{s_{eff}}\frac{1}{I_0(\kappa_{vs}a)}\int_0^a R\left(\frac{r_0}{r_b}\right)I_0(\kappa_{vs}r_0)r_0 dr_0$$

is the coupling coefficient of a relativistic electron bunch with a surface wave. Note that in the most interesting case

$$\kappa_{vs}a \equiv \frac{\omega_s a}{v_0\gamma_0} \ll 1. \quad (83)$$

the longitudinal component of the electric field of the surface wave in the vacuum channel is practically uniform over the channel cross section, and the coupling coefficient $\Gamma_s = 1$ for any transverse bunch profile.

Let us investigate the expressions for the field components of the wake surface wave in a number of limiting cases. First of all, let's consider the limiting case of a vacuum channel of small radius, when, along with inequality (83), also the condition

$$\kappa_{ds}a \ll 1.$$

is satisfied. In this case, the spectrum equation (79) can be simplified and represented in the form

$$\varepsilon(\omega) = -\frac{\kappa_d^2 a^2}{2}\left[\ln\left(\frac{1}{\nu\kappa_d a}\right) - \frac{K_0(\kappa_d b)}{I_0(\kappa_d b)}\right].$$

The roots of this equation are easy to find by the method of successive approximations. As a result, we obtain the following expression for the surface wave frequency

$$\omega_s^2 = \omega_L^2\left\{1 - \frac{\kappa_L^2 a^2}{2\varepsilon_{eff}}\left[\ln\left(\frac{1}{\nu\kappa_L a}\right) - \frac{K_0(\kappa_L b)}{I_0(\kappa_L b)}\right]\right\}. \quad (84)$$

The structure of the surface wave field has the form

$$E_z^{(s)} = E_L Z_{\parallel}(\omega_s\tau)\ln\left(\frac{1}{\nu k_L a}\right)\begin{cases}1, & r \leq a, \\ \frac{\Delta_0(k_L r, k_L b)}{\Delta_0(k_L a, k_L b)}, & r \geq a.\end{cases}$$

At $\kappa_L a \to 0$, the surface wave frequency (84) approaches to the frequency of longitudinal optical phonons. Accordingly, the structure of the wake surface wave continuously transforms into the structure of the field of longitudinal optical phonons (33) at $r_b \to 0$.

The above consideration is valid when the radius of the vacuum channel and, accordingly, the transverse size of the electron bunch is significantly less than the wavelength of longitudinal optical phonons $a \ll \lambda_L / 2\pi$, $\lambda_L = 2\pi c / \omega_L$. A more realistic case is when the dimensions of the vacuum channel and the electron bunch are of the same order of magnitude or exceed the length of the excited surface wave. We will assume that the condition

$$\kappa_d a \gg 1 \qquad (85)$$

is satisfied. Then, all the more $\kappa_d b \gg 1$. The bunch is strongly relativistic $\gamma_0^2 \gg 1$ so that requirement (83) is still satisfied. Under these conditions, the spectrum equation (79) for determination of the frequency of the wake surface wave takes the form

$$\varepsilon(\omega)\left(1 + \frac{k_0^2 a^2}{8\gamma_0^2}\right) = -\frac{\kappa_d a}{2} \tanh(\kappa_d L). \qquad (86)$$

In the case of a thick layer $\kappa_d L \gg 1$ equation (86) is simplified

$$\varepsilon(\omega) = -\frac{\kappa_d a}{2}\left(1 - \frac{k_0^2 a^2}{8\gamma_0^2}\right).$$

This equation is equivalent to the following

$$\frac{\epsilon^2(\omega)}{\epsilon(\omega) + 1} = \frac{k_0^2 a^2}{4}\left(1 - \frac{k_0^2 a^2}{4\gamma_0^2}\right), \qquad (87)$$

where $\epsilon(\omega) = -\varepsilon(\omega)$. Taking into account that the right-hand side of this equation is a large parameter, the root of equation (87) is easily found approximately

$$\omega^2 = \omega_s^2 \equiv \omega_T^2\left[1 + \Delta\varepsilon\left(\frac{4}{\kappa_T^2 a^2} + \frac{1}{\gamma_0^2}\right)\right]. \qquad (88)$$

The wake wave frequency (88) is close to the initial frequency of the surface wave (67), but slightly exceeds it. The expressions for the components of the electromagnetic surface wave in the considered limiting case have the form

$$E_z^{(s)} = E_s Z_\|(\omega_s \tau) \begin{cases} 1, & r \leq a, \\ \sqrt{\dfrac{r}{a}} \dfrac{\sinh \kappa_d(b-r)}{\sinh \kappa_d(b-a)}, & r \geq a, \end{cases}$$

$$H_\varphi^{(s)} = E_s \frac{\kappa_T a}{2} Z_\perp(\omega_s \tau) \begin{cases} \dfrac{r}{a}, & r \leq a, \\ \sqrt{\dfrac{r}{a}} \dfrac{\cosh \kappa_d(b-r)}{\cosh \kappa_d(b-a)}, & r \geq a, \end{cases}$$

$$E_s = \frac{32Q}{a^2} \frac{\Delta\varepsilon}{\kappa_T^2 a^2}. \qquad (89)$$

Behind the electron bunch $|\bar\tau| \gg 1$, the surface field has the form of a monochromatic wave. For example, for the longitudinal component of the electric field of the wake wave in the vacuum channel, we have

$$E_z^{(s)} = E_s \frac{\hat T(\Omega_s)}{\hat\tau} \cos \omega_s \tau. \qquad (90)$$

For the Gaussian longitudinal profile (40), instead of (90), we obtain

$$E_z^{(s)} = E_s e^{-\frac{\omega_T^2 t_b^2}{4}} \cos \omega_s \tau.$$

If the condition for coherent excitation of a wake surface wave by an electron bunch is satisfied $\omega_T^2 t_b^2 / 4 < 1$, the value $E_s$ is the amplitude of the longitudinal component of the electric field of the surface wave in the vacuum channel

For numerical estimates, the value of this amplitude can be represented as

$$E_s = 120 \delta_T \frac{N_0}{a^2(\mu m)} (V/cm),$$

where $N_0$ is the number of particles in the bunch,

$$\delta_T = \frac{\Delta\varepsilon}{\pi^2}\left(\frac{\lambda_T}{a}\right)^2 < 1,$$

$\lambda_T = 2\pi c / \omega_T$. So, for example, for a crystal of potassium iodide $KI$ we have $\Delta\varepsilon = 2.4$, $\lambda_T = 100\mu m$. For the radius of the vacuum channel $a = 100\mu m$, we obtain

$$E_s = 2.9 \cdot 10^{-3} N_0 \; (V/cm).$$

For $N_0 = 10^{11}$ from this formula we obtain the estimation for the electric field strength $E_L = 0.29 GV/cm$. The strength of the surface wake wave is lower than the strength of the longitudinal wake wave at frequency $\omega_L$.

Let us now consider the limiting case of a thin dielectric layer

$$\kappa_d L \ll 1. \qquad (91)$$

In this limiting case, the spectrum equation (86) is simplified and takes the form

$$\varepsilon(\omega) = -\frac{1}{2}\kappa_d^2 aL.$$

Using an explicit expression for the dielectric constant (14), this equation can be transformed to the form

$$\frac{\omega^2 - \omega_L^2}{\omega^2 - \omega_F^2} = \frac{1}{2}\frac{d_{opt}}{\varepsilon_{opt}}\frac{\omega^2 aL}{c^2}, \qquad (92)$$

The root of the quadratic equation (92), belonging to the frequency range (56) of surface waves, has the form

$$\omega_s^2 = \frac{2\omega_L^2}{1 + \alpha_T + \sqrt{(1+\alpha_T)^2 - 4\alpha_T e_0}}, \qquad (93)$$

$$e_0 = \frac{\omega_L^2}{\omega_F^2} \equiv \frac{\varepsilon_{st}}{\varepsilon_{opt}} \frac{\varepsilon_{opt} - \beta_0^{-2}}{\varepsilon_{st} - \beta_0^{-2}} < 1, \quad \alpha_T = \frac{1}{2}\frac{d_{st}}{\varepsilon_{opt}}\kappa_T^2 aL.$$

In the ultrarelativistic case $\beta_0 = 1$, we have $e_0 = {}_0$, $d_{st} = \varepsilon_{st} - 1$ and the expression for the parameter $\alpha_T$ coincides with the previously introduced parameter $\alpha_{T0}$ (69).

If the dielectric layer is so thin that the stronger condition is satisfied

$$\alpha_T \ll 1, \quad (94)$$

in comparison with (91), then the expression for the frequency of the excited surface wave follows from the formula

$$\omega_s^2 = \omega_L^2 \left[1 - (1 - e_0)\alpha_T\right] = \omega_L^2 \left(1 - \frac{\Delta\varepsilon}{2\varepsilon_{opt}^2}\kappa_T^2 aL\right). \quad (95)$$

Under these conditions, the frequency of the surface wave is close to the frequency of longitudinal optical phonons.

Behind the relativistic electron bunch in limiting case (83), the expression for the longitudinal component of the electric field (82) of the wake surface wave in the vacuum channel can be represented as

$$E_z^{(v,s)} = \frac{2Q}{a^2}\frac{1}{\Lambda_s}\frac{\hat{T}(\Omega_s)}{\hat{\tau}}\cos\omega_s\tau. \quad (96)$$

In the considered case of a thin dielectric layer the for function $\Lambda(\omega)$ we have the expression

$$\Lambda(\omega) = -\frac{\omega^2}{2(\omega^2 - \omega_L^2)}\left(\frac{\omega_L^2}{\omega^2} - \frac{\omega^2 - \omega_L^2}{\omega^2 - \omega_F^2}\right).$$

For a surface polaritons the parameter $\Lambda_s$ can be written as follows

$$\Lambda_s = \frac{1}{2(1-w_s^2)}\left(1 - \alpha_T e_0 w_s^4\right),$$

where

$$w_s^2 = \omega_s^2/\omega_L^2 = \frac{2}{1 + \alpha_T + \sqrt{(1+\alpha_T)^2 - 4\alpha_T e_0}} < 1.$$

For an electron bunch with a Gaussian longitudinal profile, the expression for the wake field (96) takes the form

$$E_z^{(v,s)} = \frac{4Q}{a^2}\frac{1-w_s^2}{1-\alpha_T e_0 w_s^4}e^{-\frac{\omega_s^2 t_b^2}{4}}\cos\omega_s\tau. \quad (97)$$

For a short bunch $\omega_L^2 t_b^2/4 \ll 1$ and a thin dielectric layer (94) instead of (97) we have

$$E_z^{(v,s)} = 2Q\kappa_T^2\frac{L}{a}\frac{\Delta\varepsilon}{\varepsilon_{opt}^2}\cos\omega_s\tau,$$

and the frequency of the surface wave (95) is close to the frequency of longitudinal optical phonons.

## CONCLUSION

The process of the excitation of Cherenkov wake electromagnetic waves by a relativistic electron bunch in an ion dielectric waveguide is studied. The ion dielectrics of the alkali-halide group are considered. In ion dielectrics of this group in the infrared and terahertz frequency ranges, there are three branches of electromagnetic waves (polaritons). Two of them correspond to transverse electromagnetic waves (transverse polaritons). In the infrared range there is also a branch corresponding to the optical longitudinal phonons (longitudinal polariton) of an ion dielectric. For all these branches, analytical expressions for the wake electromagnetic field excited by a relativistic electron bunch are obtained and studied. It is shown that in the infrared (microwave) frequency range the excited wake electric field consists of potential monochromatic wave belonging to the branch of longitudinal polaritons and a set of eigen electromagnetic waves of a dielectric waveguide (transverse polaritons).

In the presence of a vacuum channel surface polaritons appear in the spectrum of eigen waves of the ion dielectric waveguide. Surface polaritons of an ion dielectric exist in the frequency range $\omega_L > \omega > \omega_T$. Note that, in this frequency range, there are no volume transverse polaritons. The process of excitation of wake surface polaritons by a relativistic electron bunch propagating in a vacuum channel is also investigated. It is shown that relativistic electron bunches several tens of microns in sizes can serve as an effective tool for coherent excitation of accelerating wake fields in the infrared range. Ion dielectrics as plasma-like media are of considerable interest for the creation of wake dielectric accelerators of charged particles on their basis.